\documentstyle[prl,aps,psfig]{revtex}
\begin{document}
\title{Gravitational Waves from   
 Mesoscopic   Dynamics of  the  Extra Dimensions}
\author{Craig J. Hogan}
\address{Astronomy and Physics Departments, 
University of Washington,
Seattle, Washington 98195-1580}
\maketitle
\begin{abstract}
Recent models which describe our world as a brane embedded in a higher
dimensional space introduce new geometrical degrees of freedom:   
the shape and/or size of the extra dimensions, and  the position of the brane.  
These modes can be coherently excited by symmetry breaking in the early universe 
even on  ``mesoscopic'' scales as large as 1 mm, leading to  detectable
gravitational radiation. Two sources are described:  relativistic turbulence 
caused by  a first-order transition of a  radion potential, and 
Kibble excitation of Nambu-Goldstone modes of brane displacement. 
Characteristic scales and spectral properties are estimated and the prospects
for observation by LISA are discussed. Extra dimensions with scale
between 10 \AA\  and 1 mm, which enter the  3+1-D  era at cosmic temperatures 
between 1 and 1000 TeV,    produce backgrounds with energy peaked at observed 
frequencies in the LISA band, between  $10^{-1}$ and $10^{-4}$ Hz. The background 
is detectable above instrument noise and astrophysical foregrounds  if  
initial metric perturbations are excited to a fractional amplitude of $10^{-3}$
or  more, a likely outcome for the Nambu-Goldstone excitations.

\end{abstract}

\section{Mesoscopic Cosmology  of the Extra Dimensions}

 Recently
 there has been considerable  interest in models
of spacetimes with relatively large extra dimensions,
in which the familiar   
 Standard Model fields are confined to
  a   ``brane'', a 3-dimensional defect in the larger space, while gravity 
propagates in all the dimensions (the ``bulk'').  
 In some models\cite{arkanihierarchy,antoniadas,arkaniphenom}, the extra dimensions can be  
 as large as the current direct  experimental gravitational probes of the order of  a
millimeter\cite{adelberger}. The apparent (usual) Planck mass in 3+1D, $M_{Planck}$, is related
to the true  fundamental scale $M_*$ by
$M_{Planck}^2\approx M_*^2 (M_*^nb^n)$
where $b^n$ is the volume of $n$ extra large dimensions.
In another class of models (``nonfactorizable geometries'',
\cite{randallsundrum1,randallsundrum2,lykkenrandall}),
the extra dimensions can be even larger, but have   curvature $k$
  which traps gravitons in a bound state close to a  brane.
The curvature radius $b\approx k^{-1}\approx M_{Planck}^2/M_*^3$
 is again on a mesoscopic scale which may be    as large as
$\approx 1$ mm. 

The   cosmology of these models might well include radical departures from
the orderly evolution of the standard model, including episodes of violent
mesoscopic geometrical activity at relatively late times. 
New classical geometrical
   effects  remain important long after inflation, until the   Hubble
length
 $ H^{-1}\approx M_{Planck}/T^{2} $ is   larger than  the
size or curvature radius $b$ of the largest extra
dimensions\cite{arkaniphenom,csaki,rapidinflation}.
 Of particular interest   are two new geometrical degrees of freedom common to 
many of these models: ``radion''
modes controlling the size or curvature of the extra dimensions\cite{csaki,goldbergerwise},
and new Nambu-Goldstone modes corresponding to inhomogeneous  displacements
 of the brane in the extra dimensions\cite{arkaniphenom,sundrum}.
Previous treatments have focused on  microscopic 
thermal and quantum emission in these modes, but
  cosmological symmetry breaking can also create large-amplitude, coherent classical excitations 
on much larger scales  (of order $H^{-1}$)   
as the configuration of the extra dimensions and the position of the brane settle into their
present state. This highly dynamic geometric activity
generically produces a conspicuous detectable relic: an intense, classically generated
background of gravitational radiation. 
In
this paper  I explore the possibility of a direct detection
of a   
  gravitational wave background generated by coherently excited
radion and
Nambu-Goldstone modes  on the threshold of standard cosmology.

\section{Hubble Frequency and Energy Equipartition }

The main features of the background spectra can be estimated
from general scaling considerations, without reference to particular models.

\paragraph{\it Frequency:}
 The characteristic gravitational (``Hubble'') 
frequency\cite{hogangwb,maggiore}  
  redshifted to
the present day is $f_{H0}(T)\equiv H(T) a(T)/a_0$ or: 
\begin{equation}
f_{H0}=7.65\times 10^{-5}\ {\rm
Hz}\ (T/TeV)g_*^{1/6}(g_*/g_{*S})^{1/3}T_{2.728}=
 9.37\times 10^{-5}\ {\rm Hz}(H\times 1{\rm mm})^{1/2}g_*^{-1/12}(g_*/g_{*S})^{1/3}
T_{2.728}
\end{equation}
The 
estimate  is valid back to the  threshold of the extradimensional dynamics,
$H^{-1}\approx b$.
The mapping between observed frequency, 
temperature $T$ and Hubble length is shown by axis labels in
 figure 1 for frequencies in the LISA band.\footnote{Note the weak 
 dependence on the particle-physics
uncertainties encapsulated in   the  number of effective
relativistic degrees of freedom, defined in the usual way $g_*\equiv
\sum_{bosons}(T_i/T)^4g_i+ (7/8)\sum_{fermions}(T_i/T)^4g_i$ and
$g_{*S}\equiv \sum_{bosons}(T_i/T)^3g_i+
(7/8)\sum_{fermions}(T_i/T)^3g_i$). Even  above 100 GeV we expect 
  standard  relativistic cosmology derived from
relativity and thermodynamics to hold(e.g.\cite{kolb}),
$H=2.07\times 10^5\ {\rm Hz}\ g_*^{1/2}(T/GeV)^2$ 
and
$a(T)/a_0
=3.70\times 10^{-13} T_{2.728}g_{*S}^{-1/3} (GeV/T)$.}
   Extra
dimensions between $b\approx 1$ mm and $10^{-6}$ mm produce backgrounds peaked 
in the LISA band ($10^{-1}$ to $10^{-4}$ Hz);  observations with 
LIGO (up to $\approx 1000$Hz)
could detect\footnote{This requires high enough sensitivity,  $h_{rms}\approx 10^{-22}$,
to detect waves with $\Omega_{GW}\approx\Omega_{rel}$, which may  be possible with future
instrumentation.}
activity from    dimensions down to
$b\approx 10^{-14}$mm.

\paragraph{\it Amplitude:} 
 The   mechanisms considered here
 excite geometrical degrees of freedom in approximate
equipartition of energy density with the thermal relativistic matter.
A stochastic background of gravitational radiation 
with rms metric perturbations  $h_{rms}(f)$ over bandwidth $\Delta f$ contributes a fraction
of the critical density
$\Omega_{GW}(f,\Delta f=f)\approx (f/H)^2 h_{rms}^2(f, \Delta f=f)$. 
In an experiment such as LISA a stochastic background 
can only be distinguished from other sources of noise and astrophysical wave sources
by resolving the background in frequency (and to some extent, direction and polarization).
  The   rms strain produced in 
a LISA frequency resolution element is\cite{lisa98}
\begin{equation}
h_{rms}(f, \Delta f )
=4.74\times
10^{-20}f_{mHz}^{-3/2}T_{2.728}^2 h_{70}^0 (\Delta f/ 3\times 10^{-8}{\rm {\rm Hz}})^{1/2}
[\Omega_{GW}(\Delta f = f)/\Omega_{rel}]^{1/2}.
\end{equation}
The reference density is set\cite{hogangwb,maggiore} by 
the mean energy density in  all relativistic species (photons and three massless neutrinos),
$\Omega_{rel}=8.51\times 10^{-5} h_{70}^{-2} T_{2.728}^4$,
where $h_{70}$ refers to the Hubble constant.
Since the energy density of
gravitational waves redshifts like relativistic
matter, this gives a maximal bound for primordial
broad-band  backgrounds, 
  shown in figure 1 along with projected LISA sensitivity\cite{armstrong}.\footnote{A chaotic
cosmology is almost unconstrained by other data if it happens early enough. 
Scalar inhomogeneities (including entropy perturbations) from subhorizon
events earlier than about 100 GeV are   erased by neutrino and nucleon diffusion  before
nucleosynthesis begins\cite{heckler,kainulainen}, so gravitational waves may be the only direct
surviving relic.   Overproduction of black
holes or other stable relics must be avoided but these are very model-dependent\cite{carr}.
Existing data   constrain 
 gravitational waves   only  at much lower frequencies; 
above $f_{msp}=4.4\times 10^{-9}$Hz
(corresponding to $\approx f_{H0}(T\approx 60{\rm MeV}, H^{-1}\approx 453{\rm km}))$, millisecond
pulsar timing
 gives a limit at 90\% confidence\cite{thorsett},
$\Omega_{GW}< 1.14\times 10^{-4}
(f/f_{msp})^2T_{2.728}^{-4}h^0\ \Omega_{rel}.$}

\section{Turbulent   Flow  from a First-Order Radion Transition}

The radion can be a significant source of mesoscopic activity 
  if its  potential has a first-order
phase transition. The radion is stuck initially in a metastable  state with some initial
value    $b_i$ and 
thermal or quantum nucleation  leads to randomly nucleated regions
corresponding to   the final value $b_0$, accompanied by a release of internal energy.
The gravitational waves
from this  mode are easiest to describe  if for the largest extra dimension,
$b_0\le H^{-1}(T=M_*)$: the final dimensional stabilization then happens within the context of  an
approximately 3+1-dimensional cosmology.
  
On our brane this process resembles nucleation of  
bubbles or vacuum domains of Higgs scalars\cite{coleman}, with the extra complication of 
  modifications in gravity.  In an
expanding universe\cite{peierls,ignatius,kurki} 
 the bubbles 
collide and overlap,  
 creating   flows of energy with   velocities
of the order of unity. The coherence
 scale $R$ of the flows is determined  by nucleation dynamics
which    generally yields\cite{nucleation83}  $R\le \log(HT)/H\approx 10^{-2}H^{-1}$. 
A number of   models have been used to estimate 
the gravitational radiation in similar situations, from colliding bubbles and the resulting
energy  flows in the context of QCD and electroweak phase
transitions\cite{transition84,transition86,transition94,gleiser}. In the 
absence of a definite model of radion bubbles, we estimate the maximal gravitational
wave spectrum, based  on establishing a sustained relativistic turbulent cascade up to the
scale $R$ for a time $H^{-1}$.

The  
power output of a system in gravitational waves\cite{mtw,thorne}, 
  $L_{GW}\approx 0.1 L_{internal}^2/L_0$, is determined by
 the (changing quadrupolar)  flow of mass-energy $L_{internal}$, where
$L_0=c^5/G= M_{Planck}^2$ and the numerical factor $\approx 0.1$ is typical of simple
asymmetric geometries.    Flows  of 
 mass-energy  on the Hubble scale    produce  a gravitational
wave power per volume  close to $10^{-1} H^3L_0$, and integrated over
time $H^{-1}$ produce broad-band $\Omega_{GW}\approx 10^{-1}\Omega_{rel}$ at characteristic
frequency
$H$ (now shifted to $f_{H0}$).
  Flows on smaller scale $R$  create  a spectrum peaked 
at higher frequency $f_{peak}\approx f_{H0}(RH)^{-1}$ and with a smaller amplitude.  
Relativistic motions in a
volume
$R^3$ 
 with  density perturbations of order unity,  $M\propto R^3$ 
create a mass-energy flow  
 $L_{internal}\propto (M/R)\propto R^2$ which if sustained for time $H^{-1}$ gives
  $\Omega(\Delta f=f)\propto R \propto f_{peak}^{-1}$. In a narrow
band this translates to amplitude $h_{rms}(f, \Delta f= 3\times 10^{-8}{\rm {\rm Hz}})
\approx  \Omega(\Delta f=f)^{1/2}(\Delta f/f)^{1/2}(H/f)
\propto  f_{peak} ^{-2}$. To estimate the low-frequency spectrum  consider
motions of smaller velocity on the same length scale, $v\approx Rf<1$. The mass-energy flow is 
$L_{internal}\approx Mv^2 (v/R)\propto v^3$, hence the gravitational
wave power $\propto f^6$.  The maximal
tail of low-frequency waves then
has  $\Omega(\Delta f=f)\propto f^6$, hence
 $h_{rms}(f, \Delta f= 3\times 10^{-8}{\rm {\rm Hz}})
\propto f^{3/2}$. These estimates lead to the maximal spectrum shown in
figure 1  for $R,T=0.1 H^{-1},10$TeV  and $R,T=0.01 H^{-1},100$GeV.
(In the quieter case where turbulence is not sustained, the amplitude
is less than this estimate by $(RH)^{-1/2}$.)
 
\section{Gravitational Waves from Brane Displacement }

In many scenarios there is 
another   degree of freedom, the position of the brane.
Spatial inhomogeneities in this displacement correspond to 
    nearly-massless Nambu-Goldstone modes\cite{arkaniphenom,sundrum}.  
In the cosmological formation of the brane/defect, the position is in general
a random variable uncorrelated on large scales, and large-scale modes are
  excited by the  Kibble mechanism.
 Since the modes  
also represent curvature of the brane, they are efficiently coupled into
gravitational waves as viewed on the brane. This mode is generally excited 
if for the largest extra dimension, $b_0\ge H^{-1}(T=M_*)$: the 3-brane condenses as
a defect before cosmology enters the 3+1-D era.

Suppose our  brane forms at some early time as a defect
in $3+n+1$ space, spontaneously breaking the  
  Poincar{\'e} 
symmetry of the full theory. The position of the brane in each
higher dimension $j$  can be described by a field  $y_j(x_i)$ which
depends on the  coordinate $x_i$ on the brane.\footnote{The $y_j$'s are
 analogous\cite{arkaniphenom,sundrum}
to group-transformation angles $\theta $ in the Goldstone description of the pion or axion,
but have dimensions of length.  They
are related to   the canonically-normalized physical fields  $\pi$ by $y=\pi/f_b^2$ (rather than
$\theta =\pi /f_b$) where $f_b^4$ is the brane tension.   When a  $y $  is
 ``eaten'' it gets a mass
$m_y\approx f_b^2/M_{Planck}$ or $\approx (1{\rm mm})^{-1}$ for
$f_b\approx $TeV.}
 These fields  
represent propagating Nambu-Goldstone bosons, new modes in addition to the Standard Model fields
on  the brane and gravitation propagating in the bulk.  

In  a cosmological setting, the initial values of the $y_j $'s  are
not generally correlated on large scales but 
 are random for points with large separation, since the  position of the brane, when it  
condenses as a defect, is determined by local accidents.
A topologically stable 3-wall forms for example if vacua in a 4-bulk
fall into two discrete degenerate minima; the spontaneous choice of one minimum
or the other is a random variable at large separation, so the wall is not formed
initially in its (flat) ground state, but in an excited
 (wrinkled) state\cite{zeldovich,kibble}.   This 
is the same ``Kibble mechanism'' which leads in other 
contexts to     formation of cosmic strings,
 axion miniclusters, or   Goldstone excitations of scalar fields
(which are also gravitational wave sources, e.g.  \cite{hogangwb,gold82,krauss,gold95,allen}).

If the cosmology  thereby excites   these modes,   
the brane is wrinkled on all scales up to the dimensional size or curvature radius $b_0 $
(which we take to be fixed here), curving through
the extra dimensions.
Ignoring the subtleties of the  cosmological
metric and  imagining the brane in a flat spacetime, displacements directly induce an
effective 4-metric on  the brane
$g_{\mu\nu}=\eta_{\mu\nu}+h_{\mu\nu}$, where the perturbed 4-metric
is   $h_{\mu\nu}=\sum_{j=1}^n\partial_\mu y_j \partial_\nu y_j $.
Although these initial perturbations are scalars,
no symmetry prevents  transverse-traceless tensor
components\cite{mtw} from being excited dynamically for waves with $f\ge H$, creating 
classical gravitational waves.   
 Kibble excitations corresponding to   perturbations of magnitude
$\delta y $ on scale $f^{-1}$ thereby lead to  wave amplitudes of
the order of $h_{rms}(f)\approx |h^{TT}_{jk}|\approx
 | \delta yf|^2.$

In the cosmological context, the Kibble excitation is regulated by
the expansion.  
The brane location is initially
uncorrelated on large scales, so on the Hubble scale
variations  in $y$   are of order $H^{-1}$, producing waves with amplitude
$h_{rms}\approx H^2\delta y^2\approx O(1)$, comparable  in total energy  with the other forms of
energy on the brane ($\rho_{rel}\approx H^2 M_{Planck}^2$) which now  appear in $\Omega_{rel}$.
 These waves show up
today as gravitational waves at the redshifted Hubble frequency $f_{H0}(H)$.
 The maximal
spectrum corresponds to a
  background of the order of the energy density in brane fields (that is, $h=O(1)$ on each scale
as it enters the horizon), leading to  constant $\Omega(GW)$ over a range of 
higher frequencies.   The amplitude of excitation is     reduced  after
 the universe enters the classical 3+1-D era, when $H^{-1}\ge b_0$
(or after $H\le m_y$);    
variations in $y$  are at  most of order $b_0$ (or $m_y$), leading to $h_{rms}\approx (Hb_0 )^2$
   and  to a  spectrum with a corresponding $h\propto f^{5/2}$ rolloff  at lower
frequencies.  The damping scale and low-frequency spectrum contain 
information on the scale of  extra dimensions and/or the stabilization 
of the brane. The spectrum at higher frequencies is a probe of the cosmological model
during the multidimensional/preclassical period, including the excitation mechanism.

 The predicted backgrounds
can be distinguished    from other expected
signals  by their isotropy,  and  by the   distinctive 
spectra with 
rolloffs at both low and high frequencies .  Under some circumstances these 
backgrounds could be stronger
than previously contemplated  sources\cite{lisa98,thorne,allen}. 
 \acknowledgements
I am grateful for useful discussions with E. Adelberger, P. Bender, D. Kaplan, A. Nelson,
S. Sigurdsson, E. Witten, and especially G. Fuller.
I thank  
the Aspen Center for Physics, the Max-Planck-Institut f\"ur
Astrophysik,     
  the Isaac Newton Institute for Mathematical
Sciences  and the Ettore Majorana Centre for Scientific
Culture for hospitality.
This work was supported at the University of Washington
by   NASA, and at the Max-Planck-Institute f\"ur
Astrophysik by a Humboldt Research Award.
{}

\begin{figure}[htbp]
\caption{Spectra of predicted gravitational wave backgrounds in the LISA band.
The amplitude of metric strain in a frequency 
resolution element after one year of observations ($\Delta f=
3\times 10^{-8}$Hz) is plotted against frequency $f$. 
Sensitivity  limits are shown from LISA shot and acceleration noise
($1\sigma$ per resolution element after one year) and the
confusion-limited
astrophysical background  from compact white dwarf binaries (CWDB). 
The uppermost curve shows the amplitude of a   background   which has a
broad-band energy density equal to the
known cosmological  Standard Model
relativistic degrees of freedom (photons and neutrinos).
The other curves sketch   predicted backgrounds from mesoscopic activity in 
  new extra-dimensional modes. The spectrum from
relativistic flows is  
shown for ``maximal'' turbulence sustained for a Hubble time,
on scale $RH\approx 0.1 $ at $T_*=10$TeV and for $RH\approx 0.01 $ at $T_*=0.1$TeV.
There is  near degeneracy in the determination of parameters,
in particular a smaller  nucleation scale $RH$ at a lower redshift comes close to mimicing a
larger one at a higher redshift. The
spectrum   from maximally  excited  Nambu-Goldstone modes
of brane displacement is shown for the   scale-free limiting case 
 up to  the damping epoch, taken here to be 10 TeV (or $b_0=10\mu$).}
\end{figure}
\end{document}